\documentclass[aps,preprintnumbers,twocolumn]{revtex4}

\usepackage{graphicx}
\usepackage{bm}
\newcommand{\sign}{\:\!\text{sign}\:\!}
\newcommand{\mgaugino}{M_{1/2}}

\newcommand{\mgut}{M_{\text{GUT}}}

\newcommand{\mlosp}{M_{\text{LOSP}}}
\newcommand{\tb}{\tan\beta}

\newcommand{\gev}{\text{GeV}}
\newcommand{\tev}{\text{TeV}}

\newcommand{\be}{\begin{equation}}
\newcommand{\ee}{\end{equation}}
\newcommand{\etal}{{\em et al.}}

\newcommand{\ie}{{\em i.e.}}

\newcommand{\bsg}{B\to X_s \gamma}

\newcommand{\amu}{a_{\mu}}
\newcommand{\amuexp}{a_{\mu}^{\text{exp}}}
\newcommand{\amususy}{a_{\mu}^{\text{SUSY}}}
\newcommand{\amusm}{a_{\mu}^{\text{SM}}}

\newcommand{\eqref}[1]{Eq.~(\ref{#1})}

\begin{document}
% You should use BibTeX and revtex.bst for references
%\bibliographystyle{revtex}

% number  and your conference paper identification number on the
% title page in preprint mode. Multiple \preprint commands are allowed.
\preprint{
MIT--CTP--3198, 
UCI--TR--2001--31,
CERN--TH/2001--296,
hep-ph/0111004,
Snowmass P3-08}

\title{\bm{$\lowercase{g}_\mu-2$} in Supersymmetry}
% Optional argument for running titles on pages
%\title[]{}



\author{Jonathan L.~Feng}
%\email[]{jlf@mit.edu}
\affiliation{Center for Theoretical Physics,
             Massachusetts Institute of Technology,
             Cambridge, MA 02139, USA}
\affiliation{Department of Physics and Astronomy, 
             University of California, Irvine, CA 92697, USA}
\author{Konstantin T.~Matchev}
%\email[]{Konstantin.Matchev@cern.ch}
%\homepage[]{}
%\thanks{}
\affiliation{Theory Division, CERN,
             CH--1211, Geneva 23, Switzerland}

\date{October 31, 2001}

\begin{abstract}
The $2.6\sigma$ deviation in the muon's anomalous magnetic moment has
strong implications for supersymmetry.  In the most model-independent
analysis to date, we consider gaugino masses with arbitrary magnitude
and phase, and sleptons with arbitrary masses and left-right
mixings. For $\tan\beta=50$, we find that $1\sigma$ agreement requires
at least one charged superpartner with mass below 570 GeV; at
$2\sigma$, this upper bound shifts to 850 GeV. The deviation is
remarkably consistent with all constraints from colliders, dark
matter, and $b \to s \gamma$ in supergravity models, but disfavors the
characteristic gaugino mass relations of anomaly-mediation.
\end{abstract}
% \pacs{}

%\maketitle must follow title, authors, abstract and \pacs
\maketitle

%\section{Introduction}
%\label{intro}

The current world average of the muon's anomalous magnetic moment
$\amu$ differs from the standard model prediction by $2.6 \sigma$:
$\amuexp - \amusm = (43 \pm 16) \times 10^{-10}$~\cite{Brown:2001mg}.
The reported deviation is about three times larger than the standard
model's electroweak contribution~\cite{Czarnecki:2001pv}, and so
deviations of roughly this order are expected in many models motivated
by attempts to understand electroweak symmetry breaking.  Its
interpretation as supersymmetry is particularly attractive, as
supersymmetry naturally provides electroweak scale contributions that
are easily enhanced (by large $\tan\beta$) to produce deviations
larger than the standard model's electroweak corrections.  In
addition, $\amu$ is both flavor- and CP-conserving.  Thus, while the
impact of supersymmetry on other low energy observables can be highly
suppressed by scalar degeneracy or small CP-violating phases in simple
models, supersymmetric contributions to $\amu$ cannot be.  In this
sense, $\amu$ is a uniquely robust probe of supersymmetry, and an
anomaly in $\amu$ is a natural place for the effects of supersymmetry
to appear.

We begin with the most model-independent analysis possible consistent
with slepton flavor conservation. (For more details, see
Ref.~\cite{Feng:2001tr}.) In general, the reported deviation may be
explained entirely by new physics in the muon's {\em electric} dipole
moment~\cite{Feng:2001sq}.  However, this possibility is not realized
in supersymmetry, and we assume that the deviation arises solely from
contributions to $\amu$.  We then consider uncorrelated values of the
relevant SUSY parameters, allowing arbitrary gaugino mass parameters
and slepton masses and left-right mixing.  Despite the generality of
this framework, we find stringent upper bounds on charged
superpartners, with strong implications for future collider searches.
We then consider minimal supergravity, and find remarkable consistency
of the $\amu$ deviation with all present constraints from colliders,
dark matter searches, and precision observables, such as $\bsg$.  This
is not to be taken for granted: as an illustration, we show that the
current $\amu$ result disfavors anomaly-mediated supersymmetry
breaking.

Supersymmetric contributions to $\amu$ have been explored for many
years~\cite{Fayet}.  Following the recent $g_{\mu} - 2$ result, the
implications for supersymmetry have been considered in many
studies~[6-30]
%\cite{Everett:2001tq,Baltz:2001ts,Chattopadhyay:2001vx,%
%Komine:2001fz,Hisano:2001qz,Ibrahim:2001ym,Ellis:2001yu,%
%Arnowitt:2001be,Choi:2001pz,Kim:2001se,Martin:2001st,%
%Komine:2001hy,Baek:2001nz,Carvalho:2001ex,Baer:2001kn,%
%Baek:2001kh,Cerdeno:2001aj,Chacko:2001xd,Blazek:2001zm,%
%Cho:2001nf,Adhikari:2001ra,Byrne:2001yu,Komine:2001rm,%
%Chattopadhyay:2001mj,Endo:2001ym} 
in a variety of frameworks, including minimal
supergravity~\cite{Chattopadhyay:2001vx,Komine:2001fz,%
Ellis:2001yu,Arnowitt:2001be,
Martin:2001st,Baer:2001kn,Komine:2001rm}, no-scale
supergravity~\cite{Choi:2001pz,Komine:2001hy,Baer:2001kn}, models with
non-universality~\cite{Arnowitt:2001be,Cerdeno:2001aj,%
Chattopadhyay:2001mj,Endo:2001ym}, inverted
hierarchy~\cite{Baer:2001kn},
gauge-mediation~\cite{Komine:2001fz,Choi:2001pz,Martin:2001st,%
Baer:2001kn}, anomaly-mediation~\cite{Choi:2001pz,Baer:2001kn},
$R$-parity violation~\cite{Kim:2001se,Adhikari:2001ra}, and
CP-violation~\cite{Ibrahim:2001ym}, as well as flavor
models~\cite{Hisano:2001qz,Carvalho:2001ex,Baek:2001kh,%
Chacko:2001xd,Blazek:2001zm}, and string-inspired
frameworks~\cite{Choi:2001pz,Martin:2001st,Baek:2001nz,Blazek:2001zm}.

The anomalous magnetic moment of the muon is the coefficient of the
operator $\amu \frac{e}{4m_\mu} \, \bar{\mu} \sigma^{mn} \mu \,
F_{mn}$, where $\sigma^{mn} = \frac{i}{2} \left[ \gamma^m, \gamma^n
\right]$.  The supersymmetric contribution, $\amususy$, is dominated
by diagrams with neutralino-smuon and chargino-sneutrino loops.  In
the absence of significant slepton flavor violation, these diagrams
are completely determined by only seven supersymmetry parameters:
$M_1$, $M_2$, $\mu$, $\tb$, $m_{\tilde{\mu}_L}$, $m_{\tilde{\mu}_R}$,
and $A_{\mu}$.  In general, $M_1$, $M_2$, $\mu$, and $A_{\mu}$ are
complex.  However, bounds from electric dipole moments generically
require their phases to be very small.  In addition, $|\amususy|$ is
typically maximized for real parameters.  In deriving
model-independent upper bounds on superparticle masses below, we
assume real parameters, but consider all possible sign combinations;
these results are therefore valid for arbitrary phases.

\begin{figure}[tbp]
\includegraphics[width=.48\textwidth]{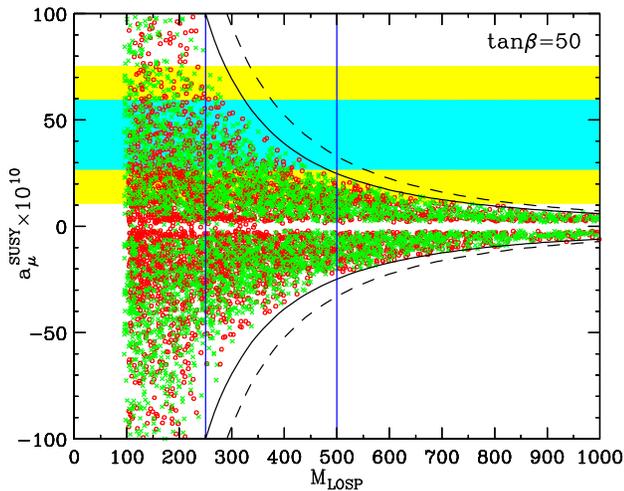}%
\caption{Allowed values of $M_{\text{LOSP}}$, the mass of the lightest
observable supersymmetric particle, and $\amususy$ from a scan of
parameter space with $M_1=M_2/2$, $A_{\mu} = 0$, and $\tb = 50$. Green
crosses (red circles) have smuons (charginos/neutralinos) as the LOSP.
A stable LSP is assumed.  Relaxing the relation $M_1=M_2/2$ leads to
the solid envelope curve, and further allowing arbitrary left-right
smuon mixing (large $A_{\mu}$) leads to the dashed curve.  The
envelope contours scale linearly with $\tb$.  The 1$\sigma$ and
2$\sigma$ allowed $\amususy$ ranges are shown, and the discovery
reaches of linear colliders with $\sqrt{s} = 500~\gev$ and 1 TeV are
given by the vertical blue lines.}
\label{fig:mnlsp}
\end{figure}

To determine the possible values of $\amususy$ without model-dependent
biases, we have calculated $\amususy$ in a series of high statistics
scans of parameter space, requiring only consistency with collider
bounds and a neutral stable lightest supersymmetric particle (LSP).
(LSPs that decay visibly in collider detectors are examined in
Ref.~\cite{Feng:2001tr}.)  We begin by scanning over the parameters
$M_2$, $\mu$, $m_{\tilde{\mu}_L}$, and $m_{\tilde{\mu}_R}$, assuming
gaugino mass unification $M_1= M_2/2$, $A_{\mu}=0$, and $\tb=50$.  The
free parameters take values up to 2.5 TeV.  The resulting values in
the $(\mlosp, \amususy)$ plane are given by the points in
Fig.~\ref{fig:mnlsp}.  We then consider arbitrary (positive and
negative) values of $M_2/M_1$, leading to possibilities bounded by the
solid curve.  Finally, we allow any $A_{\mu}$ in the interval
$[-100~\tev, 100~\tev]$, which extends the envelope curve to the
dashed contour of Fig.~\ref{fig:mnlsp}. As can be seen, allowing large
$A_{\mu}$, \ie, large left-right smuon mixing, significantly enlarges
the range of possible $\amu$. The envelope contours scale linearly
with $\tb$ to excellent approximation.

{}From Fig.~\ref{fig:mnlsp} we see that the measured deviation in
$\amu$ is in the range accessible to supersymmetric theories and is
easily explained by supersymmetric effects.

The anomaly in $\amu$ also has strong implications for the
superpartner spectrum.  Among the most important is that at least two
superpartners cannot decouple if supersymmetry is to explain the
deviation, and one of these must be charged and so observable at
colliders.  Non-vanishing $\amususy$ thus imply upper bounds on
$\mlosp$.  The {\em dashed} contour is parametrized by
\begin{eqnarray}
{}^{\rm \textstyle Arbitrary}_{\rm \textstyle LR\ mixing}: &&
\frac{\amususy}{43\times 10^{-10}} =
\frac{\tb}{50} \left( \frac{450~\gev}{\mlosp^{\text{max}}} \right)^2 
\ . \nonumber
\end{eqnarray}
If $\amususy$ is to be within 1$\sigma$ (2$\sigma$) of the measured
deviation, at least one observable superpartner must be lighter than
570 GeV (850 GeV).

These upper bounds have many implications. They improve the prospects
for observation of weakly-interacting superpartners at the Tevatron
and LHC. They also impact linear colliders: in this model-independent
framework, an observable supersymmetry signal is guaranteed at a 1.2
TeV linear collider if one demands $1\sigma$ consistency in $\amu$;
improvements in $\amu$ measurements may significantly strengthen such
conclusions.  Finally, these bounds provide fresh impetus for searches
for lepton flavor violation, which is also mediated by sleptons and
charginos/neutralinos.

We now turn to specific models, where $\amu$ is correlated to many
other observables.  We first consider the framework of minimal
supergravity, where models are completely specified by the parameters
$m_0$, $\mgaugino$, $A_0$, $\tb$, and $\sign(\mu)$.  The first three
are the universal scalar, gaugino, and trilinear coupling masses at
$\mgut \simeq 2\times 10^{16}~\gev$.  We determine the entire weak
scale superparticle spectrum through two-loop renormalization group
equations~\cite{2loop RGEs} with one-loop threshold corrections and
superparticle masses~\cite{Pierce:1997zz}.

\begin{figure}[tbp]
\includegraphics[width=.48\textwidth]{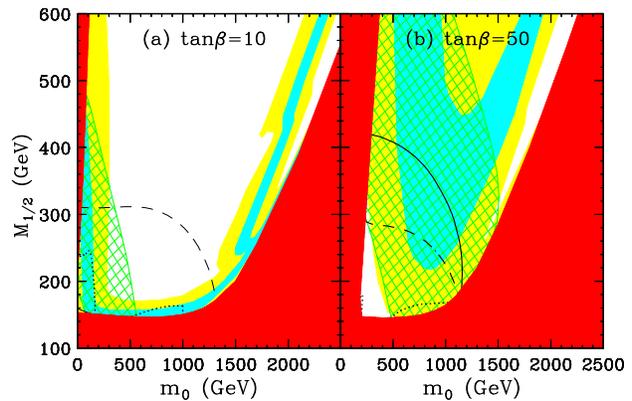}%
\caption{The 2$\sigma$ allowed region for $\amususy$ (hatched) in
minimal supergravity, for $A_0=0$, $\mu>0$, and two representative
values of $\tb$. The dark red regions are excluded by the requirement
of a neutral LSP and by the chargino mass limit of 103 GeV, and the
medium blue (light yellow) region has LSP relic density $0.1 \le
\Omega h^2 \le 0.3$ ($0.025\le \Omega h^2 \le 1$).  The area below the
solid (dashed) contour is excluded by $\bsg$ (the Higgs boson mass),
and the regions probed by the tri-lepton search at Tevatron Run II are
below the dotted contours.  }
\label{fig:amu_sugra}
\end{figure}

In minimal supergravity, many potential low-energy effects are
eliminated by scalar degeneracy.  However, $\amususy$ is not
suppressed in this way and may be large.  In this framework,
$\sign(\amususy) = \sign(\mu M_{1,2})$.  As is well-known, however,
the sign of $\mu$ also enters in the supersymmetric contributions to
$\bsg$.  Current constraints on $\bsg$ require $\mu M_3 > 0$ if $\tb$
is large. Gaugino mass unification implies $M_{1,2} M_3 > 0$, and so a
large discrepancy in $\amu$ is only possible for $\amususy > 0$, in
accord with the new measurement.

In Fig.~\ref{fig:amu_sugra}, the 2$\sigma$ allowed region for
$\amususy$ is plotted for $\mu > 0$.  Several important constraints
are also included: bounds on the neutralino relic density, the Higgs
boson mass limit $m_h > 113.5~\gev$, and the 2$\sigma$ constraint
$2.18 \times 10^{-4} < B(\bsg) < 4.10 \times 10^{-4}$.

For moderate $\tb$, much of the region preferred by $\amususy$ is
excluded by $m_h$. What remains, however, is consistent with all
experimental constraints and the requirement of supersymmetric dark
matter.  For large $\tb$, $\amususy$ favors a large allowed area that
extends to large $\mgaugino$ and $m_0 \approx 1.5~\tev$, a region of
parameter space again consistent with all constraints and possessing
the desired relic density~\cite{Feng:2000mn}.  The cosmologically
preferred regions of minimal supergravity are probed by many pre-LHC
dark matter experiments~\cite{Feng:2000gh}.  Note, however, that the
sign of $\mu$ preferred by $\amu$ implies destructive interference in
the leptonic decays of the second lightest neutralino, and so the
Tevatron search for trileptons is ineffective for $200~\gev < m_0 <
400~\gev$~\cite{Matchev:1999nb}.

We close by considering anomaly-mediated supersymmetry breaking.  One
of the most striking predictions of this framework is that the gaugino
masses are proportional to the corresponding beta function
coefficients, and so $M_{1,2} M_3 < 0$.  Anomaly-mediation therefore
most naturally predicts $\amu <
0$~\cite{Feng:2000hg,Chattopadhyay:2000ws}, in contrast to the
observed deviation.  A more detailed quantitative
analysis~\cite{Feng:2001tr} in minimal anomaly-mediation finds that,
even allowing a 1$\sigma$ deviation in $\amu$, it is barely possible
to obtain 2$\sigma$ consistency with the $\bsg$ constraint.  Minimal
anomaly mediation is therefore disfavored. The dependence of this
argument on the characteristic gaugino mass relations of anomaly
mediation suggests that similar conclusions will remain valid beyond
the minimal model.

In conclusion, the recently reported deviation in $\amu$ is easily
accommodated in supersymmetric models.  Its value provides {\em
model-independent} upper bounds on masses of observable superpartners
and already discriminates between well-motivated models.

{\em Acknowledgments} --- This work was supported in part by the
U.~S.~Department of Energy under cooperative research agreement
DF--FC02--94ER40818.

\end{document}